\documentclass[sigconf,anonymous=False,noacm]{acmart}
\setcopyright{none}
\usepackage{mdframed}
\usepackage{minted}
\usepackage{listings,subcaption}
\usepackage{tikz}
\usepackage{enumitem}

\definecolor{urlcolor}{rgb}{0.4,0.2,0.2}
\definecolor{webblue}{rgb}{0,0,.7}
\definecolor{webgreen}{rgb}{0,.5,0}
\definecolor{webbrown}{rgb}{.6,0,0}

\usepackage{tikz}
\usepackage{makecell}
\usepackage{multirow,graphicx}
\usepackage{threeparttable} 
\usepackage{booktabs}
\usepackage{diagbox}
\usepackage{amsfonts}
\usepackage{multicol}
\usepackage[ruled,linesnumbered]{algorithm2e}
\graphicspath{{figures/}} 
\usetikzlibrary{shadows}
\tikzset{every shadow/.style={opacity=1}}
\newmdenv[shadow=false,shadowcolor=black,shadowsize=0pt,linewidth=1pt,skipabove=0pt]{highlightbox}

\usepackage{pifont}
\newcommand{\cmark}{\textcolor{red}{\ding{52}}} 
\newcommand{\xmark}{\textcolor{green}{\ding{55}}}  
\newcommand{\partialcheck}{\textcolor{orange}{\ding{52}}\rotatebox[origin=c]{-9.2}{\kern-0.7em\textcolor{orange}{\ding{55}}}}

\lstdefinestyle{SQLStyle}{
  language=SQL,
  basicstyle={\footnotesize\ttfamily},
  breaklines=true,
  frame=single, 
  frameround=tttt, 
  backgroundcolor=\color{gray!5}, 
  rulesep=0pt,
 float=t,
 floatplacement=tbp,  
  numbers=none,
  keepspaces=true,
  showstringspaces=false,
  captionpos=b,
  aboveskip=1pt,
  belowskip=-20pt,
  numberstyle=\tiny\color{gray},  
  stringstyle=\color{webgreen},
  keywordstyle=\color{webblue},
  commentstyle=\color{gray},
  keywords=[2]{LATERAL, UNNEST, APPLY},
  keywordstyle=[2]\color{webblue},
  keywords=[3]{LLM, ObjColumn, ObjTable, SampleBatch, SampleNum, TableName, LimitedNum, ColumnName, ColumnType, ColumnConstraints, ColumnComment, DBName, ColumnSize},
  keywordstyle=[3]\color{webbrown},
  keywords=[4]{RAND},
  keywordstyle=[4]\color{webgreen},
}

\lstdefinestyle{prompt}{
    backgroundcolor=\color{gray!5},
    basicstyle={\footnotesize\ttfamily},
    keepspaces=true,
    showstringspaces=false,
    numbers=none,
    frame=single,
    xleftmargin=3pt,
    xrightmargin=3pt,
    aboveskip=5pt,
    belowskip=-10pt,
    captionpos=b,
    breaklines=true,             
    breakindent=-1pt,
    breakatwhitespace=true,
    tabsize=2,
    keywords=[3]{ColumnName, ColumnType, ColumnConstraints, ColumnComment, ColumnSize, },
  keywordstyle=[3]\color{webbrown},
  keywords=[4]{Samples, TableInfo},
  keywordstyle=[4]\color{webgreen},
  morecomment = [s][\color{webgreen}\bfseries]{Data samples of the current }{column}, 
}

\lstdefinestyle{prompt2}{
    backgroundcolor=\color{gray!5},
    basicstyle={\footnotesize\ttfamily},
    keepspaces=true,
    showstringspaces=false,
    numbers=none,
    frame=single,
    xleftmargin=3pt,
    xrightmargin=3pt,
    aboveskip=1pt,
    belowskip=1pt,
    captionpos=b,
    breaklines=true,             
    breakindent=-1pt,
    breakatwhitespace=true,
    tabsize=2,
    keywords=[3]{ColumnName, ColumnType, ColumnConstraints, ColumnComment, ColumnSize, k,ndv_first_k_values,min_first_k_values,max_first_k_values, most_frequent_first_k_values,first_k_values },
  keywordstyle=[3]\color{webbrown},
  keywords=[4]{Samples, TableInfo},
  keywordstyle=[4]\color{webgreen},
  morecomment = [s][\color{webgreen}\bfseries]{Data samples of the current }{column}, 
}

\lstdefinelanguage{Prompt}{
	backgroundcolor=\color{backcolour},   
	keywordstyle=\color{magenta},
	numberstyle=\tiny\color{codegray},
	basicstyle=\ttfamily\footnotesize,
	breakatwhitespace=false,         
	breaklines=true,   
    breakindent=-3.5pt,
	captionpos=b,                    
	keepspaces=true,                 
	numbers=left,                    
	numbersep=5pt,                  
	showspaces=false,                
	showstringspaces=false,
	showtabs=false,                  
	tabsize=4,
	escapeinside={`}{`},
	morecomment = [s][\color{eclipseGreen}\bfseries]{How}{?},
        morecomment = [l][\color{eclipseBlue}\bfseries]{SELECT},
        morecomment = [l][\color{eclipsePurple}\bfseries]{\$\{DATABASE_SCHEMA\}},
        morecomment = [s][\color{eclipsePurple}\bfseries]{CREATE}{;},
        morecomment = [l][\color{eclipsePurple}\bfseries]{Table},
        morecomment = [l][\color{eclipsePurple}\bfseries]{continents},
        morecomment = [l][\color{eclipsePurple}\bfseries]{countries},
        morecomment = [l][\color{codewhite}\bfseries]{\$\{TARGET_QUESTION\}},
    postbreak={
       \mbox{
           \lst@linebreakbgrd
           \rotatebox[y=0.7ex]{180}{\color{black}$\Lsh\,$}
       }
    },
}
\definecolor{pythonblue}{RGB}{0, 0, 255}
\definecolor{pythongreen}{RGB}{0, 128, 0}
\definecolor{pythonred}{RGB}{255, 0, 0}
\definecolor{pythonorange}{RGB}{255, 165, 0}
\definecolor{pythonpurple}{RGB}{128, 0, 128}

\lstdefinestyle{python}{
    language=Python,
    basicstyle={\footnotesize\ttfamily},
    backgroundcolor=\color{gray!5},   
    keywordstyle=\color{blue},        
    commentstyle=\color{pythongreen},       
    stringstyle=\color{red},          
    numbers=left,                     
    numberstyle=\tiny\color{gray},    
    stepnumber=1,                     
    numbersep=5pt,                    
    showspaces=false,                 
    showstringspaces=false,           
    showtabs=false,                   
    tabsize=4,                        
    frame=single,                      
    breaklines=true,  
    captionpos=b,   
    xleftmargin=3pt,
    xrightmargin=3pt,
}

\AtBeginDocument{%
  }

\setcopyright{acmlicensed}
\copyrightyear{2018}
\acmYear{2018}
\acmDOI{XXXXXXX.XXXXXXX}

\begin{document}

\title{ZeroCard: Cardinality Estimation with Zero Dependence on Target Databases -- No Data, No Query, No Retraining}

\author{Xianghong Xu}\orcid{0000-0003-2447-4107}
\affiliation{
  \institution{ByteDance}
  \city{Beijing}
  \country{China}
}
\email{xuxianghong@bytedance.com}

\author{Rong Kang}\orcid{0009-0005-8449-0223}
\affiliation{
  \institution{ByteDance}
  \city{Beijing}
  \country{China}
}
\email{kangrong.cn@bytedance.com}

\author{Xiao He}\orcid{0000-0001-7959-2157}
\affiliation{
  \institution{ByteDance}
  \city{Hangzhou}
  \country{China}
}
\email{xiao.hx@bytedance.com}

\author{Lei Zhang}\orcid{0009-0004-1681-1956}
\affiliation{
  \institution{ByteDance}
  \city{San Jose}
  \country{USA}
}
\email{zhanglei.michael@bytedance.com}

\author{Jianjun Chen}\orcid{0000-0002-3734-892X}
\affiliation{
  \institution{ByteDance}
  \city{San Jose}
  \country{USA}
}
\email{jianjun.chen@bytedance.com}

\author{Tieying Zhang}\orcid{0009-0003-2250-5528}
\affiliation{
  \institution{ByteDance}
  \city{San Jose}
  \country{USA}
}
\email{tieying.zhang@bytedance.com}\authornote{Tieying Zhang is the corresponding author.}

\begin{abstract}
Cardinality estimation is a fundamental task in database systems and plays a critical role in query optimization. Despite significant advances in learning-based cardinality estimation methods, most existing approaches remain difficult to generalize to new datasets due to their strong dependence on raw data or queries, thus limiting their practicality in real scenarios. To overcome these challenges, we argue that semantics in the schema may benefit cardinality estimation, and leveraging such semantics may alleviate these dependencies. To this end, we introduce ZeroCard, the first semantics-driven cardinality estimation method that can be applied without any dependence on raw data access, query logs, or retraining on the target database. Specifically, we propose to predict data distributions using schema semantics, thereby avoiding raw data dependence. Then, we introduce a query template-agnostic representation method to alleviate query dependence. Finally, we construct a large-scale query dataset derived from real-world tables and pretrain ZeroCard on it, enabling it to learn cardinality from schema semantics and predicate representations. After pretraining, ZeroCard’s parameters can be frozen and applied in an off-the-shelf manner. 
We conduct extensive experiments to demonstrate the distinct advantages of ZeroCard and show its practical applications in query optimization. Its zero-dependence property significantly facilitates deployment in real-world scenarios.
\end{abstract}

\maketitle

\section{Introduction}
Cardinality estimation aims to predict the number of tuples of SQL predicates before execution, playing a fundamental role in query optimization~\cite{zhu2024learned_survey,lan2021survey}. In recent years, learned cardinality estimators~\cite{kim2022learned,iris_lu2021pre,zeng2024price} have garnered significant attention due to their substantial performance improvements over traditional statistical methods.

\textbf{Existing works}. Most \textit{statistical} cardinality estimation methods either introduce specialized data structures (such as histogram~\cite{poosala1997selectivity,selinger1979access} and sketch~\cite{durand2003loglog,hllpp_heule2013hyperloglog,flajolet2007hyperloglog}), or adopt specific sampling strategies~\cite{haas1993fixed,chaudhuri1998random,chaudhuri2004effective} for estimation. These methods are widely applied in commercial and open-source DBMSs, like SQL Server~\cite{chaudhuri1998random,chaudhuri2004effective}, PostgreSQL~\cite{pg_cardinality}, and MySQL~\cite{mysql_cardinality}. However, their impractical assumptions about data distributions result in poor cardinality estimations~\cite{naru_yang19deep}. Afterward, learned cardinality estimation methods are developed that can significantly improve the accuracy of the estimation. In this paper, we categorize existing learned methods into four classes. (1) \textit{Data-driven} methods unsupervisedly learn the joint probability distribution of all attributes to estimate cardinality~\cite{kde_heimel2015self,kde_kiefer2017estimating,naru_yang19deep,yang2020neurocard,iam_meng2022unsupervised,DeepDB,FLAT,wu2023factorjoin,BayesCard,tzoumas2011lightweight}. 
(2) \textit{Query-driven} approaches learn supervised models to map the features of a query directly to its corresponding cardinality~\cite{dutt2020efficiently,dutt2019selectivity,kipf2018learned,nnpg_zhao2022lightweight,lpce_wang2023speeding,gl_sun2021learned,negi2023robustmscn,liu2021fauce,mscn,crn_hayek2019improved,liu2015cardinality,neuralcdf_wu2024practical}. They assume the training and test workload are independent and identically distributed (i.i.d.) and are sensitive to workload shift. 
(3) \textit{Hybrid} methods leverage both queries and data for cardinality estimation~\cite{uae_wu2021unified,zhu2021glue,li2023alece}. 
(4) In recent years, some works~\cite{iris_lu2021pre,zeng2024price,grasp_wu2025data} have attempted to reduce the dependence of applying the models to new datasets. Since they are difficult to categorize into the other three classes, we categorized them as the \textit{emerging} methods.  
For instance, Iris~\cite{iris_lu2021pre} and PRICE~\cite{zeng2024price} have attempted to employ techniques similar to pre-training in Natural Language Processing (NLP). GRASP~\cite{grasp_wu2025data} is the first method that relies entirely on query features, whereas other query-driven approaches require specific data sampling strategies.

\textbf{Challenges}. Despite substantial progress, current learned cardinality estimation methods remain impractical. They often ignore the costs associated with raw data acquisition and query log collection in real scenarios, resulting in their strong dependencies on data and queries from target databases. Even if these costs are ignored, models still must be retrained with raw data or query logs whenever they are applied to a new database. Training a dedicated model for each database is unrealistic at scale because of the high training overhead, application time constraints, and the complexity of managing a large number of models.

\textbf{Motivation}. These challenges raise a fundamental research question: \textit{Is it possible to design a single learned cardinality estimation method that does not rely on raw data or query logs?} Such an approach would eliminate the need for retraining when applied to new databases, making deployment far more practical.

\textbf{Our method}. In this paper, we introduce ZeroCard, a cardinality estimation method with zero dependence on raw data and query logs from target datasets, thereby significantly enhancing the practicality. ZeroCard is designed to primarily leverage schema semantics for cardinality estimation. Semantics, which capture the meaning of data stored in columns, can be directly obtained from the schema at negligible cost. We argue that exploiting such semantic information of the target databases may benefit cardinality estimation while eliminating the need for raw data or queries. Specifically, ZeroCard operates as follows:

\noindent(1) Semantics-driven distribution prediction. A recent study shows that using schema semantics may estimate the number of distinct values of a column without data sampling~\cite{xu2025plm4ndv}, suggesting that they may also benefit cardinality estimation. To this end, we assume that columns with specific semantics exhibit certain distributions, allowing us to capture data distributions by semantics. For instance, columns that store unique identifiers are likely to follow a uniform distribution, whereas a column that records the heights of a large population is more likely to conform to a normal distribution. By this means, ZeroCard may roughly predict the data distribution by the semantics, thus alleviating the data dependence on the new dataset. Since Pre-trained Language Models (PLMs) such as BERT~\cite{kenton2019bert} and GPT~\cite{gpt_radford2018improving} have achieved state-of-the-art (SOTA) performance across various NLP tasks, we leverage PLMs to extract the semantics of database schemas.

\noindent(2) Template-agnostic query representations. Previous methods use embedding representations for categorical attributes, which require scanning the full data to obtain the distinct values for creating learnable embeddings. This technical implementation adopted by most existing methods severely hinders model deployment on unseen datasets. We propose a unified representation method for numerical and categorical predicates to address this issue, thereby enhancing model generalization capabilities. Besides, we leverage semantics to learn latent correlations between columns, so the predicate-level modeling is agnostic to query templates.

\noindent(3) Large-scale pre-trained and off-the-shelf applicable. We pre-train ZeroCard on the task of cardinality estimation based on the GitTables~\cite{hulsebos2023gittables} dataset, which comprises large-scale relational tables. The columns in the dataset are annotated with semantic types from DBpedia~\cite{auer2007dbpedia} and Schema.org~\cite{guha2016schema}, covering thousands of distinct semantic categories. We construct a large number of queries on these tables with rich semantics, and through pre-training on this dataset, ZeroCard can perform cardinality estimation based on schema semantics. Therefore, ZeroCard applies in an off-the-shelf manner to new datasets without raw data or query dependence.

\begin{table}[t]
    \centering
    \caption{Categories of cardinality estimators and their dependencies on data and query when applying to new datasets.}
    \label{tab:compare}
    \begin{tabular}{ccccccc}
    \toprule
        Categories & Representative Works & Data & Query  \\ 
    \hline
        Statistical & histogram~\cite{poosala1997selectivity,selinger1979access} & \cmark & \xmark \\
        Data-driven  & Naru~\cite{naru_yang19deep} & \cmark & \xmark\\
        Query-driven  & MSCN~\cite{mscn} & \cmark & \cmark\\
        Hybrid   & UAE\cite{uae_wu2021unified}  & \cmark & \cmark\\
        Emerging (data) & Iris~\cite{iris_lu2021pre}, PRICE~\cite{zeng2024price} & \cmark & \xmark\\
        Emerging (query) & GRASP~\cite{grasp_wu2025data} & \xmark & \cmark\\
        Semantics-driven &  ZeroCard (ours) & \xmark  & \xmark\\
    \bottomrule
    \end{tabular}
\end{table}

\textbf{Differences and comparison}. We demonstrate representative cardinality estimators and their dependencies when applied to new datasets in Table~\ref{tab:compare}, highlighting unique advantages of ZeroCard. As shown in the table, existing methods require raw data or queries on the new dataset. While recent emerging advancements~\cite{grasp_wu2025data,iris_lu2021pre,zeng2024price} have made much progress in improving practical deployability, they still rely either on data or queries of the new datasets. To our best knowledge, ZeroCard represents the first semantics-driven cardinality estimation method that fundamentally differs from all existing learned estimators. We position this work as an exploration aimed at reducing the dependence on raw data and queries for cardinality estimation, thereby facilitating the deployment of learned cardinality estimators in practice.

In summary, we make the following contributions:
\begin{itemize}
    \item We propose ZeroCard, which leverages database semantics for cardinality estimation. To the best of our knowledge, ZeroCard is the first semantics-driven cardinality estimation method with zero dependence on raw data and query logs when applied to unseen new datasets, introducing a fundamentally new paradigm in cardinality estimation. 
    \item We develop a semantic-driven data distribution prediction method to eliminate raw data dependence and introduce a query representation approach that is agnostic to the query templates to avoid query dependence. These two semantics-based design principles distinguish ZeroCard from all previous approaches.
    \item ZeroCard is pre-trained on a large-scale dataset and is off-the-shelf applicable to new databases. We conduct extensive experiments to demonstrate the distinct contributions of our method. 
\end{itemize}

\section{Preliminary}
In this section, we formalize the cardinality estimation problem statement and review existing methods for this task. Based on this, we analyze the characteristics of these methods and identify the current challenges faced by the research community.
\subsection{Problem Statement}\label{sec:problem-state}

\noindent\textbf{Tabular Data}. Consider a relational table $T$ with $M$ attributes $A=\{a_1,a_2,\ldots,a_M\}$ and $N$ rows $T=\{t_1,t_2,\ldots,t_N\}$, where $|T|=N$ and $|A|=M$. In this paper, we consider two attribute types: \textit{numerical} (continuous-valued, e.g., integers and floating-point numbers) and \textit{categorical} (discrete-valued, e.g., strings). For numerical attributes, the value domain is a bounded interval $\textbf{dom}(a)=[l,u]\subseteq\mathbb{R}$, where $l=\inf\{v|v\in a\}$ and $u=\sup\{v|v\in a\}$. For categorical attributes, the value domain is a finite, discrete set $\textbf{dom}(a)=\{v_1,v_2,\ldots,v_D\}$, where $D$ is the number of distinct values of the attribute.

\noindent\textbf{Query}. A query represents a structured request for retrieving data from a database according to specified operations and conditions. Generally, a query consists of a conjunction of predicates, where each predicate can be formalized as ``<\textsc{attribute}> <\textsc{operator}> <\textsc{value}>'', \textsc{attribute} $=a\in A$, <\textsc{operator}> $\in\{<,>,=,\leq,\geq\}$, and \textsc{value} $\in\mathrm{dom}(a)$. In this paper, we primarily concentrate on the operators described above and ignore other operators (such as \lstinline[style=SQLStyle]{LIKE} operator) because they are not the main focus in cardinality estimation~\cite{li2023alece,han2021cardinality}.

\noindent\textbf{Cardinality Estimation}. Given a query Q, it is required to estimate its cardinality $\mathrm{Card(Q)}$ without actually executing Q. Card(Q) is the number of tuples that satisfy the conjunction of predicates. Formally, Card(Q) is equal to
\lstinline[style=SQLStyle]{SELECT COUNT(*) FROM T WHERE P}, where \lstinline[style=SQLStyle]{P} is the conjunction of predicates.

\subsection{Existing Works on Cardinality Estimation}\label{sec:related}

We briefly review existing cardinality estimation methods based on the technical paradigms.

\noindent\textbf{Statistical Methods}. Statistical methods construct various statistical features by scanning data in the target database, which can be categorized into two types: synopsis-based and sampling-based approaches. Synopsis-based methods employ specific data structures to record statistical features, with histogram~\cite{poosala1997selectivity,selinger1979access} and sketch~\cite{durand2003loglog,hllpp_heule2013hyperloglog,flajolet2007hyperloglog} being the most widely adopted structures. Sampling-based methods collect a set of samples over the target database and then execute the queries over the samples to estimate cardinality~\cite{haas1993fixed,chaudhuri1998random,chaudhuri2004effective}. Synopsis-based methods typically require scanning the entire dataset, while sampling-based approaches can alleviate this requirement. However, they still incur substantial data access costs in practice~\cite{chaudhuri2004effective}.

\noindent\textbf{Data-driven Methods}. Data-driven methods aim to learn the joint distribution of multiple attributes and subsequently estimate cardinality based on the learned distribution. They have introduced different machine learning techniques, such as auto-regression models~\cite{naru_yang19deep,yang2020neurocard,iam_meng2022unsupervised} and probabilistic graphical models~\cite{BayesCard,DeepDB,FLAT,wu2023factorjoin,tzoumas2011lightweight}, to model the distributions. For each target database, data-driven models must scan the entire dataset and train models individually.

\noindent\textbf{Query-driven Methods}. Query-driven methods extract features from queries and learn a direct mapping between query features and the corresponding cardinality. Various techniques have been introduced, such as linear model~\cite{malik2007black}, tree-based ensembles~\cite{dutt2019selectivity}, fully connected neural networks~\cite{liu2015cardinality}, multi-set convolutional network~\cite{mscn,negi2023robustmscn,kim2022learned}, etc., to capture the query features to estimate cardinality. These methods require collecting logs of queries and their cardinalities from the target database for training the model. In addition, most of them also require sample data from the database to extract specific features.

\noindent\textbf{Hybird Methods}. Hybrid methods, which model both the data distribution and query patterns, have been shown to outperform most of the data-driven or query-driven approaches~\cite{uae_wu2021unified,zhu2021glue,li2023alece}. Consequently, hybrid methods require both data scanning and the collection of query logs on the target database.

\noindent\textbf{Emerging Methods}. In recent years, several ML-based approaches have achieved significant advancements, thereby reducing the reliance of methods on the target database. Specifically, GRASP~\cite{grasp_wu2025data} represents the first data-agnostic query-based method,  enabling query-driven approaches to break free from dependence on sampled data. Iris~\cite{iris_lu2021pre} and PRICE~\cite{zeng2024price} introduce the pre-training paradigm from natural language processing (NLP) into cardinality estimation. They alleviate the models' reliance on retraining to some extent, but still depend on data access.

\noindent\textbf{Analysis}. As illustrated above, existing methods exhibit strong implicit dependencies on the target database, as detailed below:
\begin{itemize}[leftmargin=15pt]
    \item Raw data dependency. Statistical, data-driven, and hybrid methods need to scan the full data of the target database, while most query-driven ones require access to some samples in both the training and inference stages. Collectively, most existing methods implicitly require varying degrees of access to the target database for cardinality estimation.
    \item Query log dependency. Most query-driven methods assume that the training and testing workloads are independently and identically distributed (IID). Therefore, query-driven and hybrid methods require the collection of sufficiently large and comprehensive query logs.
    \item Retraining requirement. Most existing learned cardinality estimation methods necessitate retraining via data access or query log collection when deployed to new, unseen datasets.
\end{itemize}

Therefore, these implicit constraints limit their applicability in certain production scenarios.

\section{ZeroCard Overview}
\begin{figure*}[th]
    \centering
    \includegraphics[width=0.96\linewidth]{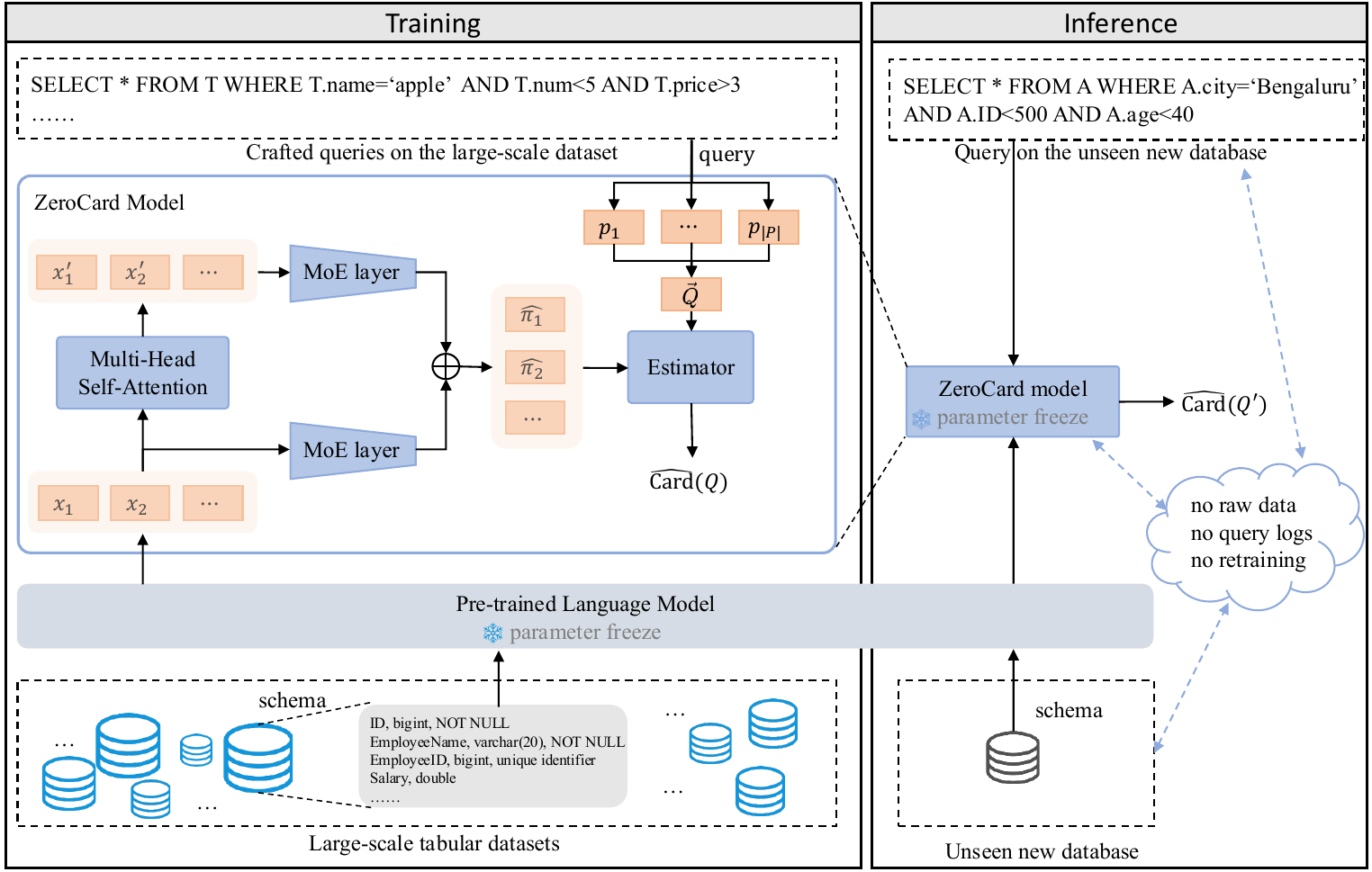}
    \caption{Architecture of ZeroCard. When pre-trained on a large-scale dataset, ZeroCard can be directly applied to unseen new databases, requiring no raw data, no query logs, or no retraining.}
    \label{fig:architecture}
    \vspace{-10pt}
\end{figure*}

In this section, we first clarify the essential concepts and assumptions of our proposed model, ZeroCard. Then we demonstrate the overall architecture of ZeroCard, with a brief illustration of its core components. 

\subsection{Assumptions}
In contrast to existing works, ZeroCard is semantics-driven, as demonstrated in Table~\ref{tab:compare}. We first clarify the concept of semantics and then state the assumptions of ZeroCard.

\noindent\textbf{Semantics}. The semantics of a database column are the meaning or the interpretation of the stored data, represented using natural language interpretations. Since PLMs demonstrate exceptional capability in natural language processing by encoding sentences into low-dimensional embedding spaces~\cite{ni2021sentencet5,reimers-2019-sentence-bert}, we leverage PLMs to encode the semantics of databases.

\noindent\textbf{Assumptions}. The proposed method, ZeroCard, operates under three foundational assumptions. (1) Meaningful columns exhibit certain kinds of distributions, making the use of models to capture data distribution reasonable. (2) The semantics of a database provide useful information for cardinality estimation, enabling the development of semantic-driven cardinality estimators. (3) The column definitions in the database schema are sufficient to capture the semantic meaning of the data, allowing for the utilization of the schema to extract semantics. (4) The maximum and minimum values for numerical columns are available, significantly alleviating dependence on raw data.

\subsection{Overall Architecture}

The overall architecture of ZeroCard is depicted in Figure~\ref{fig:architecture}, which is divided into training and inference stages, and we present the structure of ZeroCard in the training stage.

\noindent\textbf{Semantics-driven Data Distribution Prediction}. We propose to predict data distributions based on database semantics, thereby avoiding data access on the target databases. To this end, we extract semantics from the database schema by leveraging a PLM, which embeds each column into an embedding vector $x$. Since the columns in the same table may provide beneficial information, we adopt the Multi-Head Self-Attention (MHSA)~\cite{vaswani2017attention} mechanism to obtain the representation $x^\prime$ to capture the relationships between the columns. Given that columns with identical semantics may correspond to multiple potential data distributions, we design a Mixture-of-Experts (MoE) layer to enable our approach to model such relationships based on input semantics, thereby enhancing the accuracy of data distribution prediction. We model single-column and context-aware latent data distributions by the extracted semantics $x$ and $x^\prime$, respectively, and ultimately derive the column distribution $\widehat{\pi}$.

\noindent\textbf{Template-agnostic Query Representation}. We decompose queries into individual predicates and leverage semantics across columns to learn their correlations. To this end, we unify the representation of numerical and categorical predicates.

\noindent\textbf{Pre-training}. To conduct pre-training, it is necessary to collect large-scale real tabular data and then construct numerous queries. The pre-training objective encompasses two aspects: (1) data distribution prediction, which learns the implicit statistical patterns in semantics from large-scale real-world data distributions; and (2) cardinality estimation, which maps semantic information, query representations, and the estimated implicit data distributions to cardinality. By this means, the proposed model can utilize the semantics to estimate cardinality, thereby avoiding the reliance on raw data access, collection of query logs, and retraining in practice.

\noindent\textbf{Model Inference}. When ZeroCard is well pre-trained, it can perform cardinality estimation on the new database in an off-the-shelf manner, with its parameters fully frozen. Unlike most existing methods, ZeroCard requires no data access, no query log collection on the target database, and no model retraining when applying to an unseen new database.

\section{Semantics-based Distribution Prediction}

\subsection{Semantics Embedding}
The column definitions are explicitly recorded within the database schema in textual form. We first serialize the textual descriptions from the schema into token sequences, then leverage a PLM to encode these sequences into dense vector representations. Similar to~\cite{xu2025plm4ndv}, we use a naive concatenation of column definitions in the schema to obtain the column text. 

\textit{Extract textual sequences from schema}. Specifically, the definition of a column in the schema typically consists of the \textit{column name}, \textit{data type}, \textit{constraints}, and \textit{comments} in order. The textual sequence of column description is recorded as the concatenation of these four elements in order, separated by commas. Any optional elements can be omitted if undefined in the schema. Since the schema must define a name for each column, therefore, every column in the schema input can be represented as a sequence of tokens $\mathcal{T}=\{t_0,t_1,\ldots,t_{|\mathcal{T}|}\}$, where each $t_i$ is extracted from the schema.

\textit{PLM encoding}. We employ widely used PLMs to encode the textual sequences of columns to obtain their semantic representations. Specifically, we adopt sentence transformers~\cite{ni2021sentencet5,reimers-2019-sentence-bert} as our encoding methods since they are fine-tuned for the semantic textual similarity task, making them suitable for capturing column semantics. In the NLP community, numerous models are available. The encoding process can be formally expressed as follows:

\begin{align}
    \begin{aligned}
        x_i=\text{PLM}(\mathcal{T}_i),x_i\in\mathbb{R}^d,
    \end{aligned}
\end{align}
where $\mathcal{T}_i$ is the textual sequence of the $i$-th column, PLM is the encoder of  and $x_i$ is the corresponding semantic embedding.

\begin{figure}[t]
    \centering
    \includegraphics[width=0.96\linewidth]{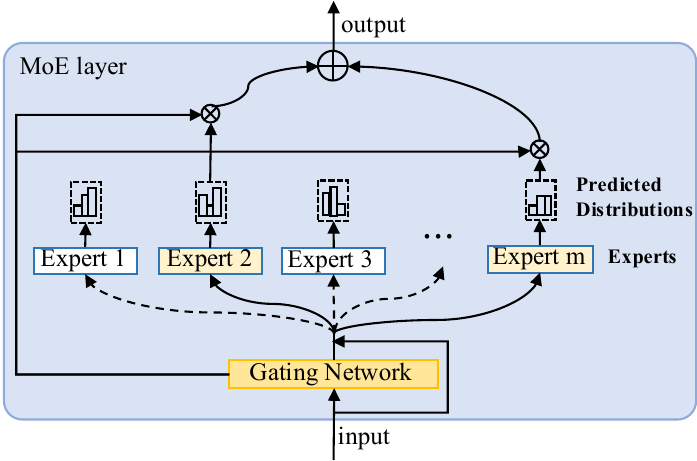}
    \caption{Structure of distribution prediction MoE layer with $k=2$. The gating network dynamically activates specific experts, each individually modeling different distributions.}
    \label{fig:moelayer}
\end{figure}

\subsection{Mixture-of-Expert Layer}
We implement the distribution prediction component based on the Mixture of Experts (MoE) architecture~\cite{garmash2016ensemble,jacobs1991adaptive,shazeer2017outrageously}, with the structure of the data distribution estimation MoE layer illustrated in Figure~\ref{fig:moelayer}. There are $m$ experts in the figure. The gating network selects appropriate experts based on the input, and the chosen experts (highlighted in yellow) are aggregated to produce the output.

\noindent\textbf{Distribution Prediction Expert Networks}. Each expert shares an identical network architecture, and we implement it using a multi-layer perceptron (MLP) to map semantic representation into the corresponding latent data distribution:

\begin{align}
    \begin{aligned}
        \mathcal{D}_i=E_i(x), \mathcal{D}_i\in\mathbb{R}^h,
    \end{aligned}
\end{align}
where $\mathcal{D}_i$ is the latent distribution produced by the $i$-th expert network $E_i$. The activation function of the MLP is ReLU~\cite{agarap2018deep}.

\noindent\textbf{Gating Network}. The gating network takes the semantic representations as input to produce an $m$-dimensional vector, and each dimension can be computed by:

\begin{align}
    \begin{aligned}
        \alpha_i=\frac{\exp\left(G(x)_i\right)}{\sum_{j=1}^m\exp\left(G(x)_j\right)},
    \end{aligned}
\end{align}
where the gating network $G(\cdot)$ is implemented using another MLP, $\alpha_i$ denotes the $i$-th dimension of the output, representing the gating weights.

\noindent\textbf{Expert Aggregation}. We employ a top-k selection mechanism where only the experts corresponding to the k highest gating weights are activated for each input. We hypothesize that meaningful columns exhibit certain distributions, which motivates the usage of the sparse expert selection strategy. This design is based on two key considerations: First, different experts may specialize in predicting particular types of distributions. Second, selecting only the most relevant experts through top-k gating is particularly appropriate based on our assumption. Specifically, the selected experts and the aggregated output are determined by:
\begin{align}
    \begin{aligned}
    \mathcal{E}&=\{i|\ |\{j:\alpha_j>\alpha_i\}|<k\}, \\
        \mathcal{D}^\prime &= \sum_{i\in\mathcal{E}}\alpha_i\mathcal{D}_i,
    \end{aligned}
\end{align}
where $\mathcal{E}$ is the set of selected expert indices and $\mathcal{D}^\prime$ is the aggregated result.

\subsection{Distribution Prediction}

\noindent\textbf{Hierarchical Distribution Fusion}. We model column data distributions from both local and global perspectives and fuse them to predict the distribution. 

\textit{Local Perspective}. The local perspective captures the intrinsic distribution patterns derived solely from the semantics of the column itself. Therefore, we can take the semantic representation of each column to obtain the local latent distribution $\mathcal{D}^\prime_{\text{local}}=\text{MoE}(x)$.

\textit{Global Perspective}. The global perspective incorporates contextual information from all predicate columns in the query to model cross-column dependencies. To effectively capture the relationships, we employ the Multi-Head Self-Attention (MHSA)~\cite{vaswani2017attention} mechanism that dynamically weights the influence of different columns based on their semantic relevance. Specifically, given a set of column representations $\{x_1,x_2,\ldots,x_n\}$. Denote $X=[x_1,x_2,\ldots,x_n]^\top$, $X^\prime=[head_1,head_2,\ldots,head_H]W^O$, each $head_i$ is an attention function:

\begin{align}
\begin{aligned}
    head_i&=\text{Attention}(XW^Q_i,XW^K_i,XW^V_i),\\
    \text{Attention}(Q,K,V)&=\text{softmax}\left(\frac{QK^\top}{\sqrt{d}}\right)V,
\end{aligned}
\end{align}
where $W^O,W^Q_i,W^K_i$, and $W^V_i$ are learnable projection matrices, and $d$ is the dimension of input vectors. This attention-based aggregation enables the model to automatically focus on the most relevant columns when constructing the global latent distribution. Then we can get $\mathcal{D}^\prime_{\text{global}}=\text{MoE}(x^\prime),x^\prime\in X^\prime$.

\textit{Predicted Distribution}. The local-global distribution modeling perspective enables comprehensive distribution prediction by combining column-specific characteristics with inter-column relationships. By simultaneously considering both local and global latent distributions, our method can produce more accurate predictions:

\begin{align}
    \widehat{\pi}=\text{softmax}(\mathcal{D}^\prime_{\text{local}}+\mathcal{D}^\prime_{\text{global}})
\end{align}
where $\widehat{\pi}$ is the predicted data distribution and $\sum_{j=1}^h\widehat{\pi}(j)=1$.

\noindent\textbf{Distribution Prediction Loss}. The objective of distribution prediction is to enable the model to approximate the distribution of columns. Therefore, the loss function of distribution prediction is the Kullback-Leibler (KL) divergence of the predicted distribution and the ground-truth distribution.

\begin{align}
    \begin{aligned}
        \mathcal{L}_{\text{dist}}&=\frac{1}{\mathcal{N}}\sum_{i=1}^{\mathcal{N}}D_{\text{KL}}\left(\widehat{\pi}_i || \pi_i\right)=\frac{1}{\mathcal{N}}\sum_{i=1}^{\mathcal{N}}\sum_{j=1}^{h}\widehat{\pi}_i(j)\log\frac{\widehat{\pi}_i(j)}{{\pi}_i(j)},
    \end{aligned}
    \label{eq:loss-dist}
\end{align}
where $\mathcal{N}$ is the number of training samples. For the $i$-th sample, $\widehat{\pi}_i$ is the predicted data distribution predicted by the model, $\pi_i$ is the ground-truth distribution, $\widehat{\pi}_i(j)$ and ${\pi}_i(j)$ are the $j$-th dimension of the $i$-th predicted distribution and ground-truth distribution, respectively.

\section{Query and Predicate Representation}
The key design principle of template-agnostic query representation is to decompose queries into predicates and learn the correlations between these predicates by semantics.
Besides, we unify numerical and categorical data representation to eliminate data-specific learnable embeddings that could limit generalization. The implementation details of this approach, including the distribution and predicate representation processes for both data types, are formally presented in Algorithm~\ref{algo:numerical-rep} and Algorithm~\ref{algo:categorical-rep}, respectively.

\begin{algorithm}[t]
\SetAlgoLined
\KwIn{Numerical predicate $\{q_l,q_u\}$, column data $a$, $l$, $u$}
\KwOut{$\pi_a,p_a$}
  $p_a \gets \vec{\mathbf{0}}_h, \pi_a\gets \vec{\mathbf{0}}_h, \vec{\mathbf{0}}_h\in\mathbb
 {R}^h$;  $|\mathcal{B}|\gets \left \lfloor \frac{u-l}{h} \right \rfloor $ ;\\
 \For{$i=1,v_i\in a;i\leq N;i\xleftarrow{}i+1$}{
 \uIf{$v_i==u$}{
 $\pi_a[h]\gets \pi_a[h]+\frac{1}{N}$; \\
 }\uElse{
 $\pi_a[\left \lfloor \frac{v_i-l}{|\mathcal{B}|} \right \rfloor+1]\gets \pi_a[\left \lfloor \frac{v_i-l}{|\mathcal{B}|} \right \rfloor+1]+\frac{1}{N}$; \\
 }
 }
 \For{$i=1,b_i\in\mathcal{B};i\leq h;i\xleftarrow{}i+1$}{
    \uIf{$q_l<b_l \wedge q_u\geq b_u$ }{
        $p_a[i]\xleftarrow{}1$; \\
    }\uElseIf{$q_u<b_l \vee q_u\geq b_u$}{
        $p_a[i]\xleftarrow{}0$; \\
    }\uElse{
        $p_a[i]\xleftarrow{} \frac{\min(q_u,b_u)-\max(q_l,b_l)}{b_u-b_l}$;
    }
}
 \textbf{return} $\pi_a,p_a$\;
 \caption{Representation of distribution and predicates for numerical columns.}\label{algo:numerical-rep}
\end{algorithm}

\subsection{Distribution Representation}
\noindent\textbf{Numerical Data}. To construct the ground truth distribution representation for numerical data, we employ a histogram-based approach with the following implementation. First, we partition the value domain range into fixed-size buckets of uniform width, and the set of buckets is denoted as $\mathcal{B}$, where the bucket boundaries are determined by the minimum and maximum values $l$ and $u$ in the column. Second, we perform frequency counting by assigning each value in the column to the corresponding bucket and accumulating the counts. Finally, we normalize the histogram by converting the absolute frequencies into relative probabilities, ensuring the sum of all bucket probabilities equals one. The detailed process is shown in Algorithm~\ref{algo:numerical-rep}. 

By this means, we can derive the distribution $\pi_a$ for column $a$. This approach preserves the original data distribution features across different columns.

\noindent\textbf{Categorical Data}. Categorical data is discrete, making it impossible to measure the ``distance'' between two values as with numerical data, which consequently makes it difficult to determine bucket sizes. To address the fundamental difference between categorical and numerical data representations, we propose to encode all discrete data into a finite domain.

We leverage a hash function $\mathcal{H}$ to transform each value in the column into an integer with a predefined value domain $\textbf{dom}(\mathcal{H})$. Specifically, we adopt MurmurHash~\cite{mmh} as the hashing function for the following critical properties: First, it has high computational efficiency that can reduce the hash overhead during data preprocessing. Second, the deterministic nature of its output ensures reproducibility across different systems and executions. Third, its superior avalanche characteristics produce uniformly distributed outputs, effectively minimizing clustering. These properties make it suitable for large-scale categorical data encoding. The representation of categorical data distributions becomes analogous to numerical data distributions when values are transformed through $\mathcal{H}$ mapping, as demonstrated in Algorithm~\ref{algo:categorical-rep}.

\begin{algorithm}[t]
\SetAlgoLined
\KwIn{Categorical predicate $\{q_a\}$, column data $a$, hash function $\mathcal{H}$}
\KwOut{$\pi_a,p_a$}
 $p_a \gets \vec{\mathbf{0}}_h, \pi_a \gets \vec{\mathbf{0}}_h,\vec{\mathbf{0}}_h\in\mathbb
 {R}^h$; $|\mathcal{B}|\gets \left \lfloor \frac{|\textbf{dom}(\mathcal{H})|}{h} \right \rfloor$;\\
 $u\gets \mathbf{sup}(\mathcal{H}), l\gets \mathbf{inf}(\mathcal{H})$ ; //boundaries.\\ 
 \For{$i=1,v_i\in a;i\leq N;i\xleftarrow{}i+1$}{
 \uIf{$\mathcal{}v_i==u$}{
 $\pi_a[h]\gets \pi_a[h]+\frac{1}{N}$; \\
 }\uElse{
 $\pi_a[\left \lfloor \frac{\mathcal{H}(v_i)-l}{|\mathcal{B}|} \right \rfloor+1]\gets \pi_a[\left \lfloor \frac{\mathcal{H}(v_i)-l}{|\mathcal{B}|} \right \rfloor+1]+\frac{1}{N}$; \\
 }
 }
 $p_a[\left \lfloor \frac{\mathcal{H}(q_a)-l}{|\mathcal{B}|} \right \rfloor]\xleftarrow{} 1$;  \\
 \textbf{return} $\pi_a,p_a$\; 
 \caption{Representation of distribution and predicates for categorical columns.}\label{algo:categorical-rep}
\end{algorithm}

\subsection{Predicate Representation}\label{sec:predicate-repre}

\noindent\textbf{Numerical Predicate}. For numerical predicates commonly used in range queries, we can analyze the overlap between the predicate interval and the corresponding data distribution to estimate the cardinality of the predicate. Specifically, given a range predicate with the query interval $\{q_l,q_u\}$, we construct a predicate vector $p_a$ with the same size as its distribution $\pi_a$. We compute each element of the predicate vector $p_a$ by evaluating the proportional overlap between the predicate interval $\{q_l,q_u\}$ and the corresponding distribution in $\pi_a$. The computation follows three cases: (1) complete containment of the bucket within the predicate interval yields a value of 1.0, (2) no overlap between the bucket and predicate interval yields 0.0, and (3) partial overlap scenarios apply linear interpolation under the uniform distribution assumption within each bucket. The procedure of constructing numerical predicate representation is shown in Algorithm~\ref{algo:numerical-rep}.

\noindent\textbf{Categorical Predicate}. For categorical predicates frequently used in equal queries, we represent them using a one-hot vector where the index corresponding to the bucket containing the hashed predicate value is set to 1. Specifically, given an equality predicate with query value $q_a$, we first compute its hash value. We then construct a predicate vector $p_a$ with the same size as its distribution $\pi_a$, where the element at the index determined by the hash value's bucket assignment is set to 1, while all other elements remain 0, as demonstrated in Algorithm~\ref{algo:categorical-rep}.


\section{Model Pre-training}

\subsection{Cardinality Estimation}

\noindent\textbf{Template-agnostic Query Representation}. 
Different queries contain varying numbers of predicates, which hinders the generation of fixed-dimensional representations required for consistent model input. To resolve this and fully realize query template-agnosticism, we first encode each predicate into a dense vector in $\mathbb{R}^h$, as illustrated in Section~\ref{sec:predicate-repre}. Additionally, each predicate has semantic representations that capture the latent correlations.
We unify these variable-length predicate vectors through an aggregation strategy, converting them into a single, fixed-dimensional query representation that enables the consistent processing of structurally diverse queries. We denote $P$ as the maximum number of predicates. For queries with fewer than$P$ predicates, we employ a padding technique to supplement the predicate sequence to length $P$, ensuring consistent input dimensionality.

\begin{align}
    \vec{Q}=\mathrm{Agg} \left (p_1||x_1^\prime,p_2||x_2^\prime,\ldots,p_{|P|}||x_{|P|}^\prime\right ), \vec{Q}\in\mathbb{R}^{d+h}
\end{align}
where $||$ is the concatenation operation, $x_i^\prime$ is the semantic information of the $i$-th predicate, and Agg denotes the predicate aggregation method, implemented using maximum pooling~\cite{boureau2010theoretical}. This approach effectively combines multiple predicate vectors into a unified query representation by selecting the most salient features across all predicates.

\noindent\textbf{Estimator}. To map the learned query representation to final cardinality predictions, we adopt an MLP as the estimation module. It takes the query representation and table size as input and is formulated as follows:

\begin{align}
    \widehat{\mathrm{Card}}(Q)=\mathrm{MLP}\left[\vec{Q}||\log N\right],
    \label{eq:estimator}
\end{align}
where the logarithmic transform of the table size helps normalize its scale, which in turn stabilizes model training.

\noindent\textbf{Cardinality Estimation Loss}. 
The loss function of cardinality estimation is shown as follows:
\begin{align}
    \mathcal{L}_\text{card}=\frac{1}{\mathcal{N}}\sum_{i=1}^{\mathcal{N}}\left(\log\widehat{\mathrm{Card}}(Q) - \log\text{Card}(Q)\right)^2.
\end{align}

\subsection{ZeroCard Pre-training}\label{sec:pretrain}
The pre-training phase, encompassing both training objectives and training data, where the latter plays a pivotal role in determining whether ZeroCard can be applied off-the-shelf. Therefore, it is essential to elaborate the data processing details in the methodology.

\noindent\textbf{Training Objective}. The training objective of ZeroCard comprises two components: data distribution prediction and cardinality estimation. We jointly optimize both tasks in an end-to-end training framework. Formally, the composite loss function is defined as:
\begin{align}
    \mathcal{L}_{\text{ZeroCard}}=\mathcal{L}_{\text{dist}}+\beta\mathcal{L}_{\mathrm{card}},
    \label{eq:loss}
\end{align}
where $\beta$ is a task-balancing hyperparameter that adjusts the relative errors during training. Given the high cost of pre-training on large-scale datasets, we do not perform exhaustive $\beta$ tuning; instead, we directly set it to 0.1 to ensure the two losses are on the same order of magnitude.

\noindent\textbf{Training Data}. We select the GitTables~\cite{hulsebos2023gittables} dataset for training our model due to its real-world practicality, comprehensive semantic coverage, and clear licensing considerations: (1) The dataset contains tables originating from authentic sources, making the trained model potentially applicable to practical scenarios. (2) It offers comprehensive semantic coverage, containing approximately 1 million tables with over 2 thousand annotated semantic types, enabling robust learning of diverse data patterns. (3) All tables are licensed. We exclusively use MIT-licensed tables, permitting unrestricted usage of the trained ZeroCard model in both academic research and commercial applications, facilitating model distribution to the broader community.

Specifically, we first exclude columns lacking meaningful names, those containing either single-character identifiers or purely numerical representations such as numeric sequences, scientific notation values, and timestamp-formatted strings. Then, we randomly generate 1-8 predicate queries per table and record the ground truth cardinality for each query by scanning the original table. Finally, we filter out the query with selectivity greater than 0.9, as such generated predicates may lack sufficient features in the training stage. Finally, we generated 2,918,000 queries for training.

\section{Experiments}

\begin{table}[t]
    \centering
    \caption{Data distribution statistics of the train/test sets, where ``Card.'' indicates the ground truth cardinality and \# Queries represents the number of queries.}
    \begin{tabular}{c|cccccccc|c|c|cccc}
\toprule
    (\# Queries)  &   & Mean & 50-th & 90-th  & 99-th \\
\hline
    \multirow{2}{*}{\makecell{Train \\ (2,918,000)}} & Card. & 38.91 & 3 & 89 & 453 \\
                           & N & 236.32 & 149 & 402 & 1336\\
\hline
    \multirow{2}{*}{\makecell{Test \\ (2,384,156)}} & Card. & 2,776.36  & 117 & 2,525 & 50,381\\
                          & N  & 15467.85 & 999 & 22,371  & 281,285 \\
\bottomrule
    \end{tabular}
    \label{tab:stats}
\end{table}

\subsection{Experimental Setups}\label{sec:exp-setup}
\noindent\textbf{Dataset Preparation}. We select GitTables~\cite{hulsebos2023gittables} as our training dataset, as elaborated in Section~\ref{sec:pretrain}. To avoid data leaks and achieve a comprehensive evaluation, we deliberately select another large-scale dataset originating from the real world, TabLib-sample~\cite{eggert2023tablib,tablib-v1-sample}, as our test set. We employ the same preprocessing and query generation methods on it and obtain a test set containing 2,384,156 queries. We demonstrate the cardinality and the table size distribution of the train/test sets in Table~\ref{tab:stats}. It is obvious that the table sizes and ground truth cardinality of queries of the test set are orders of magnitude larger than those in the training set. It indicates that the data distribution of the test set differs significantly from that of the training set, and it has more complex scenarios and is sufficiently comprehensive.

\begin{table*}[t]
    \centering
    \caption{Overall performance on the large-scale test dataset containing 2,384,156 test cases.}
    \begin{tabular}{cc|ccccccc|cccccc}
\toprule
         & Failure & \multicolumn{7}{c|}{Estimation Error}  &  Inference  & Model\\
    Method & Rate (\%) & Mean & Max & 50\% & 75\%& 90\% & 95\% & 99\% & Time (s) & Size \\
\hline    
    Histogram (AVI) & 10.79 & 3.13 & 61953.50 & 1.12 & 1.75 & 3.91 & 7.36 & 31.15 & 5,382.56 & 255,395.68 MB\\
    Histogram (EBO) & 5.16 & 2.77 & 25748.27 & 1.16 & 1.68 & 3.01 & 5.47 & 20.80 & 5,378.09 & 255,395.68 MB\\
    Histogram (MinSel) & 0  & 5.46 & 39069.18 & 1.20 & 2.08 & 5.30 & 10.98 & 60.00 & 5,359.19 & 255,395.68 MB\\
    Sampling (1\%) & 40.72 & 1.83 & 100.00 & 1.21 & 1.55 & 2.26 & 3.17 & 10.00 & 1,044.43 & 5,339.97 MB\\  

\hline
    ZeroCard & 0 & 12.23 & 28172.84 & 2.39 & 6.11 & 20.31 & 42.50 & 162.11 &  622.66 & 28.18 MB\\
\bottomrule
    \end{tabular}
    \label{tab:mainexp}
\end{table*}

\noindent\textbf{Competition Methods}. ZeroCard is the only approach that requires zero data or queries from the test dataset. Since no existing learned cardinality estimators can be performed in an off-the-shelf manner like ZeroCard, directly comparing them with ZeroCard would be somewhat unfair. Thus, we first construct some heuristic baselines derived from statistical methods.

\textit{Sampling}. We uniformly sample a fraction $r$ (where $0<r\leq 1$) of all tuples from the table at random. To estimate the cardinality of a new query, we evaluate its selectivity on the sampled data and scale the result by the sampling rate $r$.

\textit{Histogram}. Modern DBMSs typically employ per-column histograms with various heuristics for cardinality estimation~\cite{dutt2019selectivity}. 
We construct the histogram of each column with at most 200 buckets and obtain the selectivity of each predicate by the corresponding histogram. We compute the query selectivity using the following heuristics: (1) Attribute Value Independence (AVI) assumes the columns are independent, and the query selectivity is the product of each column. The AVI heuristic is widely adopted by many commercial and open-sourced DBMSs such as PostgreSQL~\cite{pg_avi}, Oracle~\cite{oracle_avi}, and SQL Server~\cite{sqlserver_card}. (2) Exponential BackOff (EBO) heuristic is used in SQL Server~\cite{sqlserver_card}, which is designed to use the 4 most selective predicates with diminishing impact to compute the query selectivity. EBO is formulated as $s_q=s_{(1)}\times s^{\frac{1}{2}}_{(2)}\times s^{\frac{1}{4}}_{(3)}\times s^{\frac{1}{8}}_{(4)}$, where $s_q$ is the selectivity of the query, $s_{(i)}$ is the $i$-th most selective fraction across all column predicates. (3) Another heuristic is also adopted by SQL Server~\cite{sqlserver_minsel}, which uses the minimal selectivity across all predicates, denoted as MinSel.

We construct the histogram for each column only once to avoid the additional resource consumption caused by repeated creation.

\noindent\textbf{Evaluation Criteria}. We conduct a comprehensive evaluation of cardinality estimators across four critical aspects: estimation accuracy, failure, time efficiency, and storage overhead.

\noindent(1) \textit{Estimation accuracy}. We measured the accuracy by the widely used metric q-error~\cite{q_error_moerkotte2009preventing}:
\begin{align}
    \mathrm{q-error}=\max\left(\frac{\mathrm{Card}(Q)}{\widehat{\mathrm{Card}}(Q)}, \frac{\widehat{\mathrm{Card}}(Q)}{\mathrm{Card}(Q)}\right),
\end{align}
where $\mathrm{Card}(Q)$ is the groundtruth cardinality and $\widehat{\mathrm{Card}}(Q)$ is the estimated cardinality. 

\noindent(2) \textit{Failure}. We define estimation failure as the model producing an estimated value of 0. This scenario has rarely been explicitly discussed in prior works, yet its significance for both evaluation completeness and practical deployment cannot be overlooked. For evaluation, such failure cases render q-error mathematically undefined (due to division by zero), so they are often omitted from analysis, leading to an incomplete assessment. For real-world DBMS deployment, estimation failure may cause severe issues in query optimization. Specifically, an estimated cardinality of 0 may be misinterpreted as ``no matching rows'' in systems like MySQL, thereby triggering the skipping of certain plans during the optimization process (e.g., skipping index selection or adopting flawed join order decisions)~\cite{mysql_skip}. Though simple checks can be implemented to handle cases where the estimated cardinality is 0 (e.g., set the cardinality of 0 to 1), this measure merely serves as a post-processing step and fails to address the root cause of estimation failure itself.

\noindent(3) We evaluate \textit{time efficiency} through the inference time required when applying the method to new datasets, including any necessary schema analysis, data or query collection, model adaptation, feature extraction, and estimation.

\noindent(4) \textit{Model size} represents the required memory for building models to perform cardinality estimation.

\subsection{Overall Performance on Large-scale Datasets}

The performance of baselines and the proposed method on large-scale datasets is shown in Table~\ref{tab:mainexp}, wherein the q-error quantiles os performed after excluding failure cases. Therefore, the quantile-based q-error metrics of different methods are not directly comparable because the excluded failure cases vary across methods. Based on the result in Table~\ref{tab:mainexp}, we can get the following observations:

\begin{itemize}[leftmargin=15pt]
    \item In histogram-based methods, the approach utilizing the MinSel heuristic outperforms those adopting the AVI and EBO heuristics. AVI and EBO involve the multiplication of selectivities across multiple columns, which may cause the query's selectivity to drop below the representable range of 32-bit floating-point numbers, resulting in estimation failure. Histogram-based methods require the longest inference time and the largest storage space. This is because for unseen new datasets, they need to scan the data to construct histograms, with their time and space consumption being proportional to the volume of the data.
    \item The sampling-based method exhibits the highest failure rate, with 40.72\% of cases failing on the test set, which aligns with practical experience. Specifically, sampling a small volume of data may result in no samples meeting the query conditions, while increasing the sample size would compromise efficiency. 
    \item On the large-scale evaluation, ZeroCard exhibits no cardinality estimation failures. However, when compared to the histogram using the MinSel heuristic, another approach with zero failure cases, ZeroCard demonstrates higher q-error across all quantiles. This indicates that, from the sole perspective of estimation accuracy, ZeroCard, as a semantics-driven cardinality estimation method, fails to deliver strong performance, as it cannot outperform statistical methods.
\end{itemize}

While ZeroCard appears to have the highest q-error among all methods, an in-depth analysis yields the following insights about ZeroCard, especially considering that ZeroCard has no access to raw data on the test set.

\begin{itemize}[leftmargin=15pt]
    \item Exceptional inference efficiency. ZeroCard takes only 622.66s to finish the inference on the large-scale dataset, including extracting semantics from the schema via PLM. Its inference efficiency is significantly lower than that of other methods.
    \item ZeroCard maintains a fixed model size. Different from statistical methods, the number of parameters in the proposed model remains constant, independent of the volume of raw data.
    \item No methods can consistently outperform ZeroCard. On the one hand, ZeroCard and Histogram (MinSel) are the only two approaches without estimation failures, whereas all other methods exhibit failure cases. On the other hand, the failure cases can be treated as instances with extremely large q-error values, so the q-error across all quantiles of the other three methods increases accordingly. Specifically, the 90th percentile q-error of the Histogram (AVI) method becomes significantly large (due to 10.79\% of its cases having extremely large errors), surpassing ZeroCard's 90th percentile q-error of 20.31; similarly, the sampling-based method, with 40.72\% failure cases, makes its 75th percentile q-error exceed that of ZeroCard either. In addition, the maximum estimation error of ZeroCard is less than that of all histogram-based methods.
\end{itemize}

Despite the overall estimation accuracy of ZeroCard being lower than the best-performing histogram-based method, the q-error increase at the 99th percentile is no more than 3 times. Considering ZeroCard has zero dependency on the raw data of the unseen dataset, its estimation accuracy is relatively reasonable.

\subsection{Ablation Study}\label{sec:ablation}

\begin{figure}[t]
    \centering
    \includegraphics[width=0.96\linewidth]{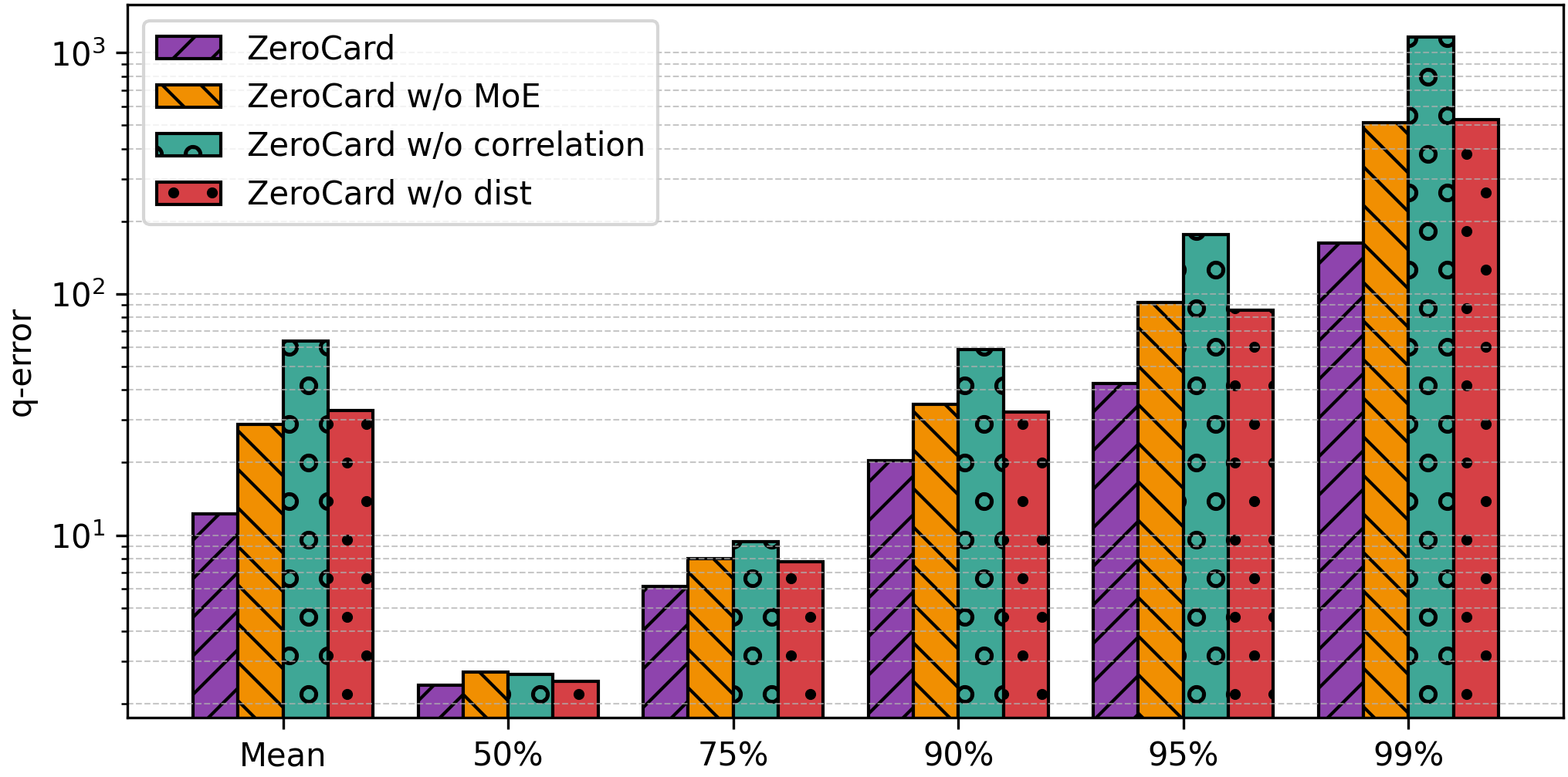}
    \caption{Ablation study: performance of ZeroCard and its variants.}
    \label{fig:ablation}
    \vspace{-10pt}
\end{figure}

The effectiveness of our method stems from the joint effects of multiple modules. To investigate the contributions of each module, we develop the following variants of ZeroCard: \textbf{w/o MoE} represents replacing the MoE layer of the model with a vanilla MLP to predict the data distribution, \textbf{w/o correlation} denotes removing the semantic correlation representations in the estimator in Equation~(\ref{eq:estimator}), and \textbf{w/o dist} indicates eliminating the distribution prediction procedure. The performance of ZeroCard and its variants is depicted in Figure~\ref{fig:ablation}, where it is evident that removing any module leads to a significant increase in estimation error. The analysis of the effectiveness of each module is presented below.

\noindent\textbf{Mixture of distribution prediction experts}. Replacing the MoE module results in the mean q-error of ZeroCard to more than double, which indicates that the design of the MoE layer in ZeroCard enables more beneficial data distribution predictions. Specifically, replacing the MoE layer with a vanilla MLP to predict the data distribution is equivalent to using only one expert. It is relatively difficult for a single distribution prediction model to predict the distributions, as similar semantics may correspond to different data distributions, and the predicted distribution may be distorted. In contrast, the MoE architecture can handle this scenario far better than a vanilla MLP model.

\noindent\textbf{Column correlation}. Removing the column correlation component significantly increases the estimation error: the mean q-error is 63.85 (over five times higher than that of ZeroCard), and the 99th percentile q-error exceeds 1162. Additionally, ZeroCard w/o correlation variant model performs substantially worse than other models across all metrics except the 50th percentile q-error. On the one hand, the performance of ZeroCard w/o correlation has a minor drop that may be attributed to the randomly constructed SQL queries. Some SQL queries involve columns with no inherent correlations; thus, failing to capture such non-existent correlations has little impact on performance. On the other hand, capturing the semantic correlations between columns is crucial for ZeroCard, as it enables the model to learn beneficial features for cardinality estimation among the potential correlated columns.

\noindent\textbf{Distribution prediction objective}. Eliminating the data distribution prediction component makes ZeroCard directly take the semantics and query representation as input to estimate cardinality. Without the auxiliary of the predicted distribution, the performance of ZeroCard w/o dist declines, and its effectiveness is similar to that of ZeroCard w/o MoE in most percentiles. It validates that if we replace the MoE layer with a vanilla model and enforce it to predict data distributions for all required semantics, using the predicted potential distorted distribution is comparable to that of the model without data distribution prediction in most cases. However, ZeroCard w/o dist performs worse than ZeroCard w/o MoE in mean q-error. Given that there are over two million test cases, this higher mean error indicates that ZeroCard w/o dist derives extremely poor estimation in certain scenarios. It demonstrates the necessity and importance of the distribution prediction objective.

\subsection{Impact of Hyperparameters} 
\begin{figure}[t]
    \centering
    \includegraphics[width=0.96\linewidth]{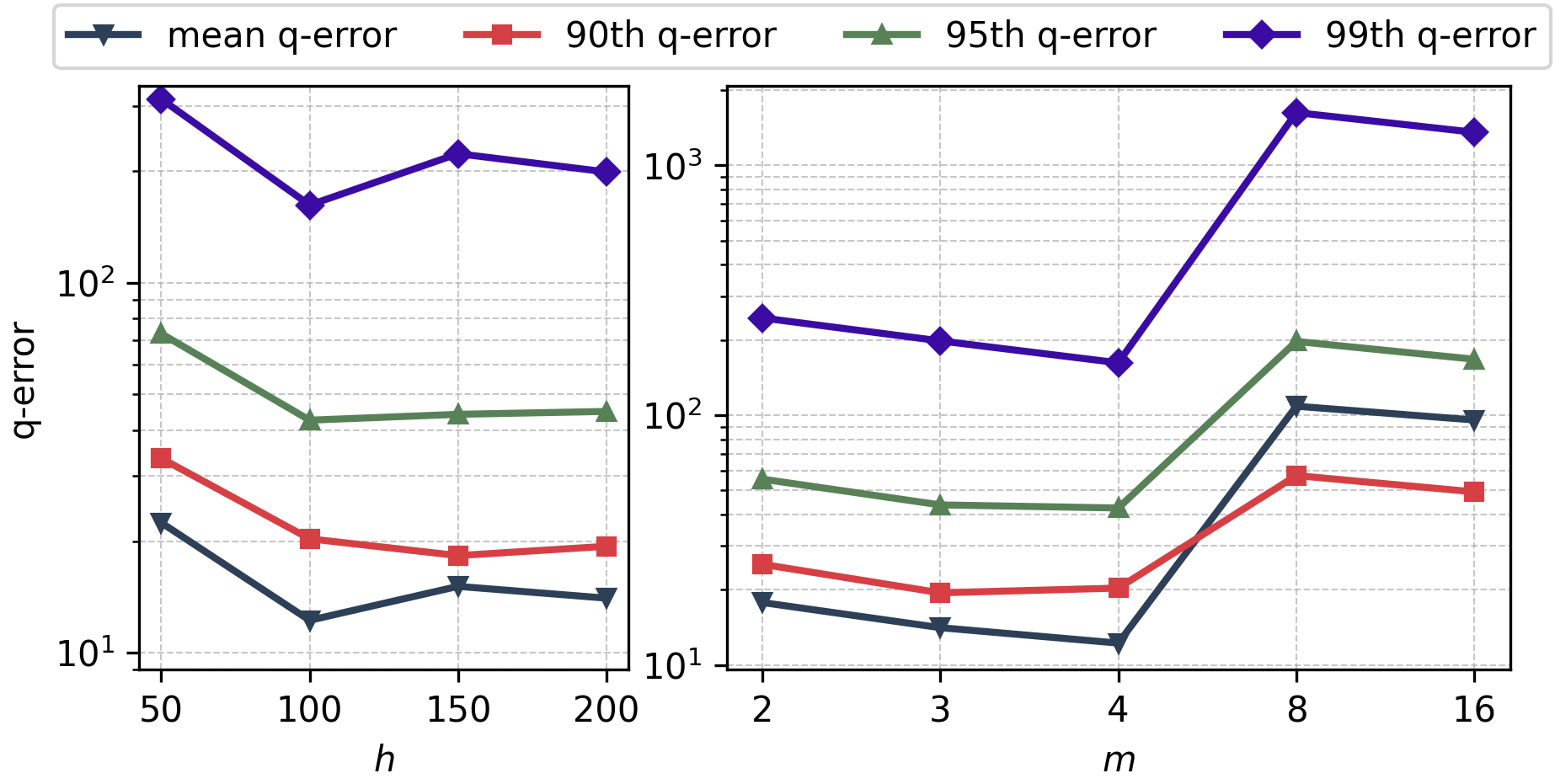}
    \caption{Performance of ZeroCard under different hyperparameter settings, where $h$ is the data distribution dimension size and $m$ is the number of experts.}
    \label{fig:hyper-h}
    \vspace{-10pt}
\end{figure}

There are two crucial hyperparameters in ZeroCard: the dimensions of distribution representation $h$ and the number of experts $m$ in MoE. We investigate the impact of these hyperparameters, and the performance is depicted in Figure~\ref{fig:hyper-h}. We present the mean, 90th, 95th, and 99th q-error, as their numerical differences are observable in a logarithmic scale.

\noindent\textbf{The dimensions of distribution representation}. We investigate the distribution representation dimension size in \{50, 100, 150, 200\}. Two key findings are derived from Figure~\ref{fig:hyper-h}, first, increasing $h$ from 50 to 100 leads to a significant performance improvement across all metrics. In contrast, when $h$ is further increased from 100 to 150 and 200, only minimal improvements are observed in a few metrics, while the 99th percentile q-error rises substantially. This indicates that the distribution dimension $h$ affects model performance: a small dimension limits the representation ability of the model, while overly large dimensions bring little benefit and even worsen the tail performance in certain scenarios. Additionally, increasing $h$ will increase the model's overhead. Therefore, we set the data distribution size $h$ as 100 in ZeroCard.

\noindent\textbf{The number of experts}. We search the number of experts $m$ in \{2, 3, 4, 8, 16\}. We observe that the model’s performance with increasing $m$ follows a pattern similar to that of increasing $h$: when $m$ is increased to 4, the estimation error of ZeroCard declines in most metrics. However, further increasing $m$ to a value greater than 4 leads to a rise in estimation error compared to that of $m=4$. Similar to increasing the dimension size $h$, increasing the number of experts $m$ also introduces additional overhead. This issue, which arises from increasing $m$, may be caused by the model overfitting to the training set. In fact, as $m$ increases, the performance of the model on the training set gradually improves. Because there is a significant difference between the data distributions of the training and test sets, consistent with the statistics presented in Table~\ref{tab:stats} (Section~\ref{sec:exp-setup}). Therefore, we set $m$ to 4 in ZeroCard for better generalization.

In summary, ZeroCard is not particularly sensitive to hyperparameters, including the distribution size $h$ and the number of experts $m$. We intentionally use training and test sets with significantly different data distributions, and the performance remains largely unchanged according to different hyperparameters. This confirms that ZeroCard maintains robust generalization.

\subsection{Efficiency Analysis}\label{sec:efficiency}

\noindent\textbf{Time efficiency}. For training time efficiency, our model requires training on large-scale datasets, and it takes approximately 8 hours to train on a single NVIDIA A100 GPU. While this training time is relatively long, the model no longer needs parameter updates once fully trained. For the inference time efficiency, the time consumption for semantic embedding generation is 597.29s on the test set. However, once these semantic embeddings are obtained, the subsequent inference time becomes negligible, requiring only 25.37s during the model inference stage. In the inference stage, the average inference time per query is about 0.3ms due to batch parallelization.

\noindent\textbf{Space efficiency}. ZeroCard eliminates the requirement of accessing raw data for synopsis creation or data sampling, so it needs no additional space beyond the model parameters. The proposed model has a fixed storage requirement of less than 30 MB, which is independent of the raw data and queries. Therefore, the space efficiency of our approach is practical for real-world deployment.

\subsection{Comparison to Learned Estimators}
\noindent\textbf{Baseline Selection}.
Existing learned cardinality estimators typically require collecting raw data or query logs on the target dataset, while ZeroCard is completely different from the previous works, making comparisons with existing learned methods somewhat unfair. Nevertheless, we selected some representative methods for comparison.

Methods that access the raw data gain direct knowledge of the distribution of the target databases, so it is expected that they may significantly outperform ZeroCard. Consequently, comparing ZeroCard with these methods may yield limited insights. In addition, most query-driven cardinality estimation methods require sampling. On the one hand, modifying their code to remove this dependency is non-trivial and may alter the original models. On the other hand, the number of tables in our experiments is several orders of magnitude larger than those in previous studies, and re-implementing all competing methods would be computationally prohibitive. For these reasons, we focus our comparison on the following representative query-driven cardinality estimation methods. 
 
\begin{itemize}[leftmargin=15pt]
    \item MSCN~\cite{mscn}. MSCN is a representative query-driven method.  We follow the hyperparameter settings specified in the original paper but omit the bitmap construction to avoid reliance on sample data.
    \item NeuroCDF~\cite{neuralcdf_wu2024practical}. NeuroCDF leverages neural networks to predict the underlying cumulative distribution functions for cardinality estimation.
    \item GRASP~\cite{grasp_wu2025data}. GRASP is the first data-agnostic query-driven cardinality estimation method.
\end{itemize}

\noindent\textbf{Settings of Dataset}. Since query-driven methods typically perform poorly on OOD queries, we re-partitioned the test set containing 2,384,156 queries to train them to avoid evaluating them on OOD queries. The number of training epochs is set to 100. Specifically, for each query template in the dataset, we randomly reserve 20\% of the queries for testing and use the remaining queries to train the query-driven methods. In total, we allocated 775,650 queries for testing and the rest for training. Model parameters of ZeroCard remain frozen throughout the experiment.

\begin{figure}[t]
    \centering
    \includegraphics[width=0.9\linewidth]{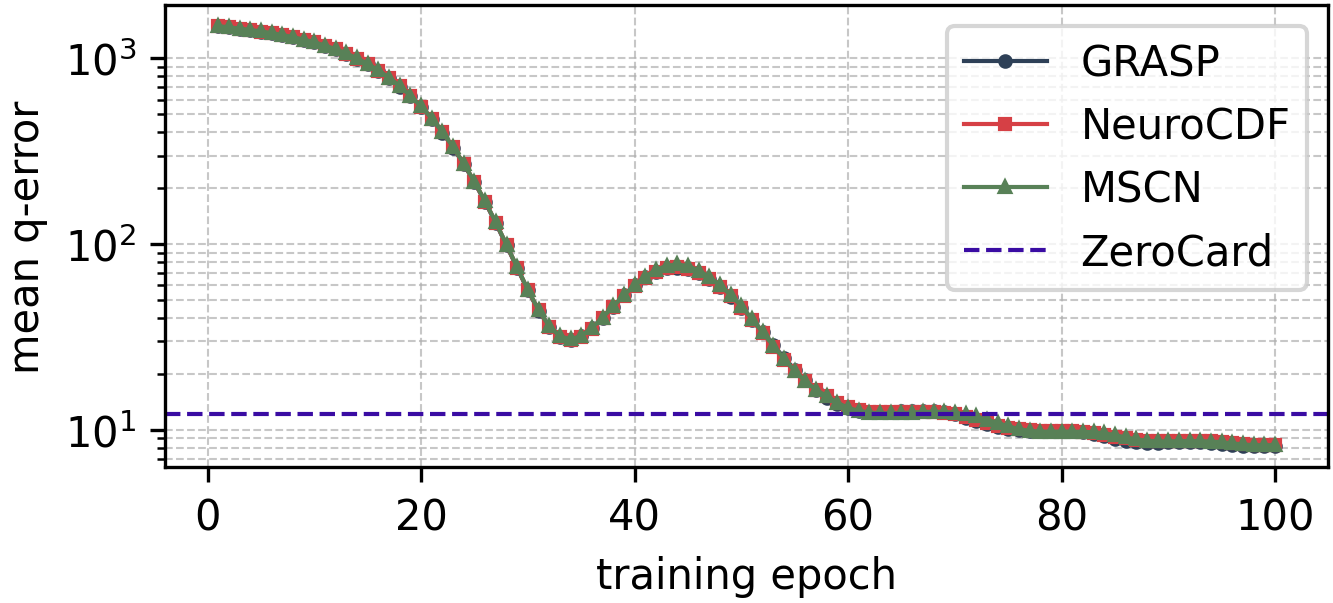}
    \caption{Performance of query-driven methods across training epochs, compared with ZeroCard.}
    \label{fig:query-driven}
\end{figure}

\noindent\textbf{Results and Analysis}. We compared the performance of the query-driven methods across training epochs with that of ZeroCard, as shown in Figure~\ref{fig:query-driven}. Based on the results presented in the figure, we draw the following conclusions:

    \textbf{\underline{(1)}} As the number of training epochs increases, though there is a temporary uptick, the query-driven methods exhibit an overall decrease in q-error, and they eventually converge to the range of 8.18–8.33. Furthermore, the performance differences among these three query-driven baselines are not significant when converged. This suggests that query-driven methods require sufficient training before being applied.
    \textbf{\underline{(2)}} The performance of ZeroCard is shown in the figure. All query-driven methods require more than 70 training epochs to outperform ZeroCard. This suggests that when the training time budget is limited, query-driven methods may not outperform ZeroCard significantly.
    \textbf{\underline{(3)}} Query-driven methods require training before deployment, whereas ZeroCard eliminates the need for retraining. Training these three query-driven methods on large-scale tables consumed approximately 117 A100 GPU-hours, which is substantially greater than that of pretraining ZeroCard. Additionally, query-driven methods struggle to handle OOD queries, making them inadequate for addressing ad hoc queries in practice.

\section{Application Analysis}
ZeroCard is the first semantics-driven cardinality estimation method that eliminates dependencies on raw data, query logs, and retraining for applying to unseen new databases, yet it only supports cardinality estimation for single-table queries. Therefore, this naturally leads us to the question: What contributions can ZeroCard offer the research community, given that it is limited to single-table query applications?

\subsection{Constraints of Application Scenarios}

While numerous learned cardinality estimation methods have been proposed in the research community, very few have been applied in practical systems, primarily due to three key dependencies illustrated in Section~\ref{sec:related}. As a result, in practical applications, many companies typically adopt sampling-based methods for cardinality estimation~\cite{han2024bytecard,idxl_peng2023data,idxl_peng2024online,chaudhuri2004effective,chaudhuri1998random}, and they must make trade-offs between sample size and estimation accuracy. In summary, most existing learned cardinality estimation methods require modifications when applied to query optimization in practical scenarios, largely due to their various dependencies. Additionally, these methods often involve compromises that lead to reduced accuracy.

\subsection{ZeroCard for Query Optimization}

The proposed method has demonstrated promising practicality in cardinality estimation on a large-scale real-world dataset; however, its effectiveness in query optimization remains unknown. Since integrating cardinality estimation methods into the query optimizer is non-trivial, we demonstrate the application of ZeroCard in the index recommendation task~\cite{wu2024automatic} to validate its utility for query optimization. Index recommendation is a query optimization technique widely offered by many commercial entities (such as Oracle~\cite{chakkappenautomatic}, Meta~\cite{yadav2023aim}, Microsoft Azure~\cite{das2019automatically}, and Meituan~\cite{idxl_peng2023data,idxl_peng2024online}), and the cardinality estimation results directly influence the quality of index recommendations.

\noindent\textbf{Task Definition of Index Recommendation}. Given a dataset $\mathbf{D}$ and a workload $\mathcal{W}$, the task of index recommendation is to develop an advisor $f$ to produce a set of indexes $\mathcal{I}$, to reduce the execution latency of $\mathcal{W}$. To avoid the overhead of physically creating indexes, the ``what-if'' analysis technique~\cite{chaudhuri1998autoadmin} is typically used to evaluate the cost reduction of candidate indexes. In index recommendation, ``what-if'' analysis typically relies on estimated statistics, including cardinality, to calculate the query cost under different given index sets. Therefore, variations of cardinality estimation results can lead to differences in costs, which in turn affect the final recommended index set $\mathcal{I}$.

\noindent\textbf{Experimental Settings}. For simplicity, we adopt a widely used heuristic index recommendation algorithm~\cite{yadav2023aim} as the index advisor $f$. We leverage VIDEX~\cite{kang2025videx} as the ``what-if'' optimizer of MySQL, which facilitates the injection of cardinality and efficient evaluation of index costs. We use two widely used open-source datasets, namely TPC-H~\cite{tpch_poess2000new} and Join Order Benchmark (JOB)~\cite{JOB_leis2015good}, with all experiments conducted on MySQL 5.7. JOB has 113 queries, which we used without modification to ensure fairness and facilitate reproducibility. TPC-H provides 22 SQL query templates; we use the scale factor of 1 and generate 22 corresponding queries using the official query generation program. Neither JOB nor TPC-H data is included in the training set of ZeroCard, so there is no data leak in this experiment. For the recommended index sets by different cardinality estimation methods, we created the indexes, executed the corresponding workload, and recorded the latency. To avoid the impact of randomness and cache, we repeated the execution of each query five times and reported the median. We compare ZeroCard with the following methods:

    \textbf{(1)} EBO (full). We use the full data of each column to construct histograms and use the EBO heuristic for cardinality estimation.
    \textbf{(2)} EBO (sample). We refer~\cite{chaudhuri1998random} to construct histograms by randomly sampling for cardinality estimation.
    \textbf{(3)} MySQL. We use the default cardinality estimated by MySQL.

\begin{figure}[t]
    \centering
    \includegraphics[width=0.96\linewidth]{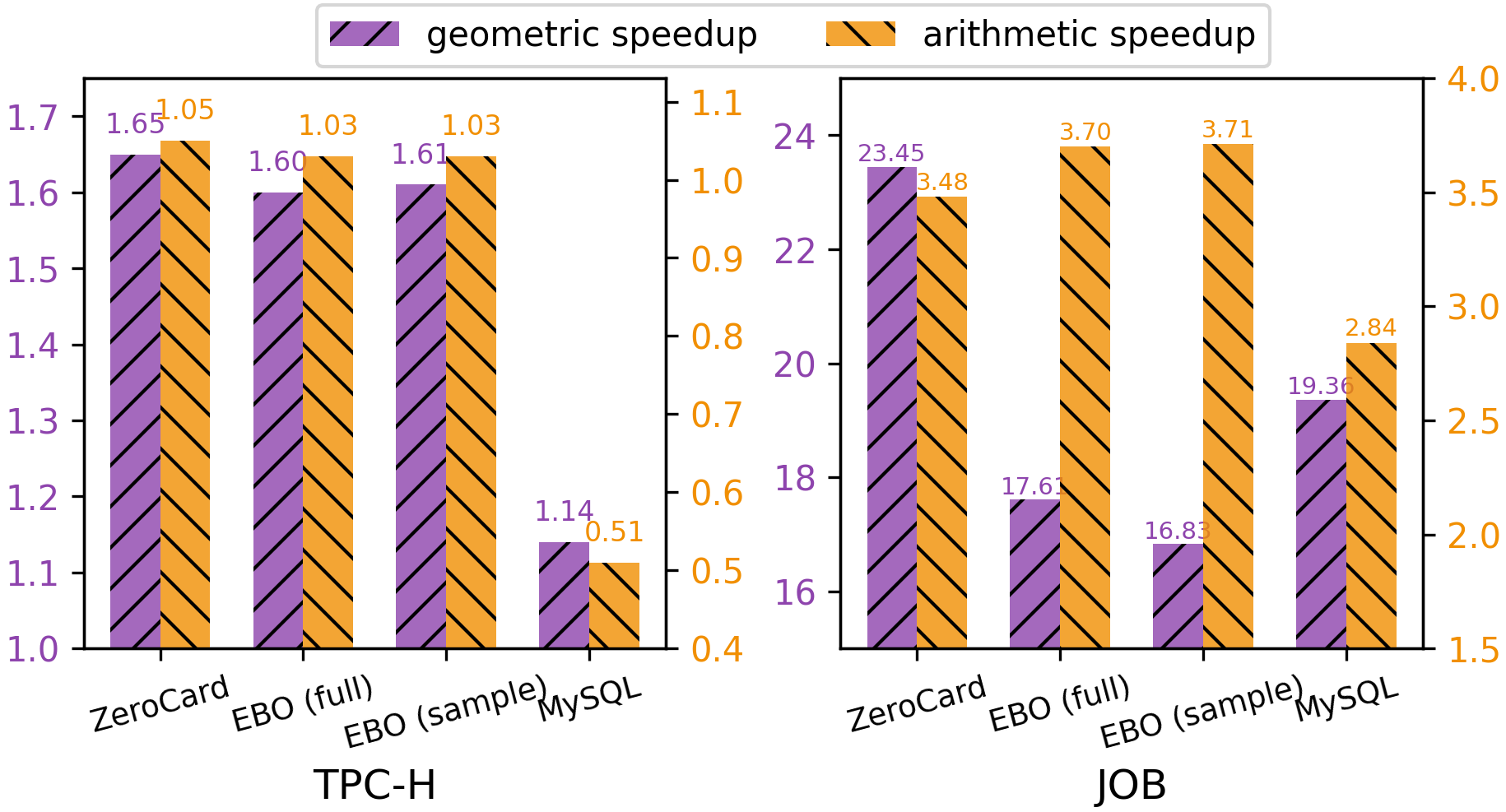}
    \caption{Workload speedup comparison of different cardinality estimation methods for index recommendation on TPC-H and JOB.}
    \label{fig:index-perf}
    \vspace{-15pt}
\end{figure}

\noindent\textbf{Performance and Analysis}. Since the arithmetic mean speedup is prone to being dominated by long-running queries, the geometric mean better reflects the speedup of short queries. We report both arithmetic and geometric speedups, with the latency without indexes used as the baseline for their calculation. The performance of these cardinality estimation methods for index recommendation is depicted in Figure~\ref{fig:index-perf}, and we can draw the following conclusions:

   \textbf{\underline{(1)}} Across both benchmarks, ZeroCard achieves better geometric speedups than all other methods. However, in terms of arithmetic speedup, it performs slightly worse than the histogram-based methods on JOB. Notably, ZeroCard outperforms MySQL’s default cardinality estimation results across all metrics. This observation, on one hand, validates the finding from prior studies~\cite{flow_loss,lee2023analyzing} that improvements or declines in cardinality estimation accuracy do not necessarily translate to corresponding improvements or declines in query performance. A possible reason is that optimizers select query plans from a set of candidate plans primarily based on relatively lower cost, rather than the absolute values of cost or cardinality.
    On the other hand, it demonstrates that despite exhibiting relatively large cardinality estimation errors, ZeroCard retains promising practical value for index recommendation, a key query optimization task.
    \textbf{\underline{(2)}} EBO (full) solely achieves better geometric speedup than EBO (sample) on JOB, while the performance difference between them is negligible in other scenarios. This result validates that although the sampling-based histogram construction strategy introduces errors, its overhead advantage makes it acceptable for practical use.
    \textbf{\underline{(3)}} MySQL’s default cardinality estimation achieves better geometric speedup than the histogram-based methods on JOB. However, the indexes recommended by MySQL’s default cardinality estimation degrade query performance, which is attributed to performance regressions in relatively long-running queries. This finding demonstrates that the cardinality estimation methods currently widely adopted in open-source DBMS still hold substantial potential for improvement.

Experiments on the index recommendation task show that ZeroCard offers unique practical value. It can be directly applied in real-world scenarios without accessing raw data or collecting query logs, and it yields promising results.


\section{Limitations} 
Although the proposed method has demonstrated distinct advantages in estimating cardinality with minimal dependency on unseen new datasets, it has some limitations.

\noindent\textbf{Semantics determine estimation quality}. ZeroCard's performance depends on correct semantic alignment between the actual column data and its schema-defined semantics. If wrong semantics are provided, ZeroCard will produce inaccurate estimations. Besides, it may struggle to treat artificially designed data distributions.

However, on the one hand, organizations can restrict arbitrary naming conventions through code reviews to alleviate misalignment issues. 
On the other hand, advanced techniques in the column annotation task~\cite{suhara2022annotating} can generate appropriate names for columns. Additionally, in recent years, products from commercial companies (such as Databricks~\cite{databricks_comments} and Amazon AWS~\cite{aws_enrich_metadata}) have attempted to automatically enrich the schema descriptions.

\noindent\textbf{Single-table predicates applicable}. ZeroCard is designed for cardinality estimation of SQL queries involving single tables. This restriction arises because the corpus used for pre-training ZeroCard does not have primary and foreign key relationships, resulting in a focus on single-table data. 

Nonetheless, single-table cardinality estimation models are still acceptable in practical DBMS systems. For instance, MySQL optimizes all queries (including both single-table and multi-table joins) by relying solely on single-table cardinality estimation.

\noindent\textbf{Maximum number of predicates}. We fixed the maximum number of predicates of a query on a table as 8, and the experiments and conclusions in this paper are based on queries with 1 to 8 predicates. It may not be feasible to use ZeroCard if the number of predicates in a single table exceeds 8.

In spite of this, we analyzed over 2.36 million SQL templates repeatedly executed within an anonymous commercial company, finding that queries with more than 8 predicates in a single table are rare (less than 0.9\%) in practical scenarios. Thus, setting the maximum number of predicates to 8 is sufficient to handle over 99\% of use cases.

\noindent\textbf{Query types}. Our method supports limited query types, as illustrated in Section~\ref{sec:problem-state}. Specifically, ZeroCard cannot estimate cardinality for queries involving \lstinline[style=SQLStyle]{LIKE} operation or aggregation operations such as \lstinline[style=SQLStyle]{GROUP BY} and \lstinline[style=SQLStyle]{HAVING}.

That said, \lstinline[style=SQLStyle]{LIKE} operation is not the primary focus in cardinality estimation. Furthermore, some practical DBMS (e.g., MySQL) do not require cardinality estimation for aggregation operations. Therefore, ZeroCard's lack of support for these cases does not compromise its utility.

\noindent\textbf{Inference efficiency}. ZeroCard has an extremely low efficiency when only CPU resources are available and needs to compute semantic embeddings, with the calculation taking about 1s per column. This may not meet the requirements of time-constrained scenarios.

Although ZeroCard's efficiency is low in the worst-case scenario, this situation is uncommon in practice. An organization has a limited number of database schemas, making it feasible to precompute the semantics embeddings for all schemas and store them. When the semantic embedding is cached, the estimation time of ZeroCard becomes trivial, even when running on CPUs.

\section{Conclusions}
In this paper, we introduce ZeroCard, a well-pretrained, off-the-shelf cardinality estimation model that operates with zero dependency on raw data, query logs, or retraining on target datasets. ZeroCard is the first semantics-driven cardinality estimation method that employs the MoE architecture and a unified predicate representation to generalize across schemas. We pretrained ZeroCard on a dataset with 2.9 million queries from hundreds of thousands of tables. Extensive experiments on cardinality estimation and query optimization demonstrate the distinct contributions of ZeroCard to the community. Furthermore, ZeroCard will be open-sourced to foster the development of semantics-driven cardinality estimation and advance query optimization research.

In future work, we will explore using data augmentation techniques to expand ZeroCard’s training dataset to improve its overall performance. Furthermore, we plan to investigate join relationships in tabular data to train semantics-driven cardinality estimators that support joins and more operators. Additionally, we aim to extend ZeroCard’s practicality, such as replacing MySQL’s native cardinality estimator with ZeroCard and applying it to large-scale commercial databases for query optimization.


\bibliographystyle{ACM-Reference-Format}
\bibliography{sample,ref}

\end{document}